\documentstyle[aps,preprint]{revtex}

\draft

\begin{document}

\title{Analytical results for a hole in an antiferromagnet}

\author{Y. M. Li,$^{a,b}$  N. d'Ambrumenil,${^a}$  L. Yu,${^{c,d}}$
and Z. B. Su ${^d}$ }

\vspace{0.2cm}
\address{a) Department of Physics, University of Warwick,
        Coventry CV4 7AL, United Kingdom\\
b) Department of Physics, Qufu Normal University, Shandong, China\\
c) Intern. Center for Theoretical Physics, P.O.Box 586, 34100 Trieste, Italy\\
d) Institute of Theoretical Physics, Academia Sinica, Beijing 100080, China}

\maketitle

\begin{abstract}
The Green's function for a hole moving in an antiferromagnet
is derived analytically in the long-wavelength limit.
We find that the infrared divergence is
eliminated in two and higher dimensions so that
the quasiparticle weight is finite.  Our results also suggest that
the hole motion is polaronic in nature with a bandwidth proportional to
$t/J \exp [-c (t/J)^2] $  ($c$ is a constant).
The connection of the long-wavelength approximation to the first-order
approximation in the cumulant expansion is also clarified. 
\end{abstract}

\bigskip
\bigskip




It has been recognised for some time \cite{review} that
understanding the motion of a hole in a 2D antiferromagnet (AF)
would be an important first step towards a complete
understanding of the effect of doping on
the CuO$_2$ planes of high T$_c$ cuprates, which
are known to show antiferromagnetism in the undoped
case. The AF with one hole is also a highly non-trivial
correlated electron system, and is therefore of fundamental
interest from a purely theoretical point of view.
There have already been many studies of the one-hole problem
including those based on exact diagonalizations (ED) of
small clusters \cite{review,PZSD,Leung,CS,Dagotto},
the self-consistent Born approximation \cite{SVR,KLR,LM,Martinez,Marsiglio},
the restricted basis method \cite{restricted},
the Bogoliubov-de Gennes (BdeG) equation \cite{Su},
and classical descriptions \cite{BNK,SS}
(see \onlinecite{review} for further references).

The few analytical results which do exist have proved very
valuable. However,
most of the previous studies have involved numerical calculations (even 
the studies using the SCBA have to solve Dyson's equation numerically 
for small clusters). This has left a few points which still
need clarification. We mention two of these. Firstly, 
in the small $J/t$ limit the SCBA yielded a
power law dependence for the hole band width [$\sim t(J/t)^\alpha$]
\cite{KLR}.
This is consistent with the results of ED's on small clusters
and gives support to the `string' picture \cite{review}. Numerical
calculations based on the BdeG equation
\cite{Su} and another variational approach \cite{CS} have suggested that
the motion of the hole is polaronic in nature in a wide parameter region.
However, both of the approaches use a
Born-Oppenheimer type approximation,
which explicitly breaks translational invariance, and it is not clear
how much this may have affected the conclusions. Secondly, 
although numerical calculations on clusters
show that the hole has a finite quasiparticle weight,
there is still some uncertainty as to
whether the quasiparticle weight vanishes or not  
in the thermodynamic limit \cite{QPW}.

Here, by treating  spin waves in the long-wavelength (continuum) limit, we
derive an analytical expression for the hole
Green's function for the hole momentum close to the  band minimum 
($\pi/2,\pi/2$).
As we work directly in momentum space, translational invariance
is {\it not} broken. This allows us to confirm 
explicitly the polaronic behaviour
of the hole.
The analytical expression we obtain for the  hole Green's function
can also be used to directly examine the hole quasiparticle weight.
Our expression for the Green's function 
shows that the infrared catastrophe,
which leads to the vanishing of the quasi-particle weight in the 1D case,
is eliminated in 2D (and higher dimensions),
so that there is a finite quasiparticle weight in 2D. This is
consistent with many other studies.
We also show that our approach is equivalent to a cumulant expansion 
and is suitable for large and
intermediate $J/t$, as for the usual polaronic problem.  Our approach
is therefore complementary to the SCBA, 
which is better in the small $J/t$ limit.

Our study is based on the $t-J$ model.  Treating the spin waves as the
collective excitations of the Heisenberg antiferromagnet,
the following  effective Hamiltonian  for the $t-J$ model
has been obtained by previous authors \cite{SVR,KLR}
\begin{eqnarray}
\tilde{H}_1 & = & H_0 + V ,
\nonumber \\
H_0 & = &   \sum_q \omega_q \beta_q^\dagger \beta_q,
\label{tJ}  \\
V & = &  \frac{tz}{\sqrt{N}} 
  \sum_{kq} h_{k-q}^\dagger h_{k} [u_q \gamma_{k-q} + v_q \gamma_{k})
          \beta_q^\dagger +(u_q \gamma_{k} +v_q \gamma_{k-q})\beta_{-q} ].
 \nonumber
\end{eqnarray}
Here ${h_k}$ and $\beta_q$ are the annihilation operators of the hole and
the spin wave,
$z$ is  the coordination number ($z=4$ for a  2D square lattice),
$\gamma_q=\sum_\delta e^{i q \cdot \delta}/z$
with $\delta$ the unit vectors to nearest neighbors, 
and $N$ is the number of the lattice sites.
The spin wave excitation spectrum
$\omega_q =Jzs \nu_q$ with $\nu_q=\sqrt{1-\gamma_q^2}$ and  $s=1/2$.
The Bogoliubov transformation coefficients are
$u_q=[(1+\nu_q)/(2 \nu_q)]^{1/2}$ and $v_q=-{\rm sgn} (\gamma_q)
(u_q^2-1)^{1/2}$.
Although  (\ref{tJ}) is not an exact mapping
of the $t-J$ model,
very good  agreement between the
results obtained from the effective Hamiltonian and
those from the original $t-J$ model  have been demonstrated
for small clusters by many authors \cite{LM,Martinez,Marsiglio}.
We take the Hamiltonian (1) as our starting point.

The hole Green's function  is defined as:
\begin{equation}
G(k,\bar{t}) =-i \langle  T h_k^H (\bar{t}) h_k^{H \dagger} (0)
      \rangle,
\end{equation}
where $\bar{t}$ denotes time throughout the paper.  $h_k^H (\bar{t})$ is
the Heisenberg operator with respect to $H$.  The thermal average 
$ \langle \cdots \rangle$ is for the spin subsystem.  
Since there is no hole for the spin subsystem, we can write
\begin{eqnarray}
G(k,\bar{t})& = &-i\theta(\bar{t})
     \langle  T  e^{i H_0 \bar{t} } h_{k}
              e^{-i H_0 \bar{t} }
     e^{-i \int_0^{\bar{t}} d\bar{t_1}     V (\bar{t_1})    }
   h_{k}^\dagger  \rangle,
\nonumber \\
& \equiv &  -i \theta(\bar{t}) \sum_{m=0}^\infty
 G_{m}(k,\bar{t}),
\label{Gexp}
\end{eqnarray}
where
\begin{eqnarray}
G_{m}(k,\bar{t}) &=& \frac{i^{2m}}{(2m)!}
        \int_0^{\bar{t}} d\bar{t}_1 \dots \int_0^{\bar{t}} d \bar{t}_{2m}
       \langle  T h_k (\bar{t}) V(\bar{t}_1) \cdots V(\bar{t}_{2m})
      h_k^\dagger  \rangle.
\nonumber \\
& = &  \frac{(itz)^{2m}}{(2m)!N^m} 
      \sum_{k_1  q_1, \dots, k_{2m} q_{2m}}
        \int_0^{\bar{t}} d\bar{t}_1 \dots \int_0^{\bar{t}} d \bar{t}_{2m}
   \langle  T M_{k_{2m}, q_{2m}}(\bar{t}_{2m}) \cdots
    M_{k_1,q_1} (\bar{t}_1)   \rangle
\nonumber \\
  &&  \langle 0 | T h_{k}
    (\bar{t}) \rho_{k_{2m}, q_{2m}}(\bar{t}_{2m}) \cdots
    \rho_{k_1,q_1} (\bar{t}_1)   h_{k}^\dagger(0) | 0 \rangle.
\label{G2m}
\end{eqnarray}
Here $\rho_{k,q} (\bar{t})=h_{k-q}^\dagger (\bar{t}) h_k(\bar{t})$, and
\begin{equation}
M_{k,q} (\bar{t}) = (u_q \gamma_{k-q} + v_q \gamma_{k})
   \beta_q^\dagger(\bar{t}) +(u_q \gamma_{k}
  +v_q \gamma_{k-q})\beta_{-q}(\bar{t}) .
\end{equation}
Formally we are treating $V$ as a perturbation. The operators $O(\bar{t})$
are now defined in an interaction picture with: $O(\bar{t}) =$
$\exp(i H_0 \bar{t}) O \exp(-i H_0\bar{t})$.

The hole part in (\ref{G2m}), $ \langle 0 |  h_{k}
    (\bar{t}) \rho_{k_{2m}, q_{2m}}(\bar{t}_{2m}) \cdots
    \rho_{k_1,q_1} (\bar{t}_1)   h_{k}^\dagger(0) | 0 \rangle$,
   equals
   $\langle 0|  h_{k}  (\bar{t})
    h_{k}^\dagger(0) | 0 \rangle$  when
$k_i=k-\sum_{l=1}^{i-1} q_l$, and is zero otherwise. We can therefore
trace out the hole part and write (\ref{G2m}) into
\begin{eqnarray}
G_{m}(k,\bar{t}) &=&\frac{(itz)^{2m}}{N^m(2m)!}  (2m)!
        \int_0^{\bar{t}} d\bar{t}_{2m}  \dots \int_0^{\bar{t}_3} d
\bar{t}_{2}
       \int_0^{\bar{t}_2} d \bar{t}_1
\nonumber \\
 && \sum_{q_1 \dots q_{2m}} \langle 
   M_{k-\sum_{i=1}^{2m-1} q_i, q_{2m}}(\bar{t}_{2m}) \cdots
    M_{k-q_1,q_2}(\bar{t}_2) M_{k,q_1} (\bar{t}_1) \rangle.
\label{Gexp2}
\end{eqnarray}
In general, it is  impossible to obtain an analytical
expression for (\ref{Gexp2}) for large $m$
since the momentum  $k_i$ in $M_{k_i, q_i} (\bar{t}_i)$ varies
with time $\bar{t}_i$, reflecting the ``history'' of the
distortion of the spin background induced by the hole.

Here we derive  an analytical expression for the Green's function
treating spin waves in the long wavelength or continuum limit.
Based on Wick's theorem, the expectation value (see (\ref{Gexp2})),
$\langle  M_{k_{2m}, q_{2m}}(\bar{t}_{2m}) \cdots
     M_{k,q_1} (\bar{t}_1) \rangle$, 
is composed of contractions like
\begin{eqnarray}
B_{k_i k_jq_iq_j} (\bar{t}_i,\bar{t}_j)
&=& \langle   M_{k_i, q_i}(\bar{t}_i)
    M_{k_j,q_j} (\bar{t}_j)  \rangle
\nonumber  \\
&=& C_{k_ik_jq_iq_j}  \delta_{q_i,-q_j}
      [  \langle  \beta_{q_i}(\bar{t}_i)
          \beta_{q_i}^\dagger (\bar{t}_j) \rangle
       + \langle  \beta_{q_i} ^\dagger (\bar{t}_i)
          \beta_{q_i} (\bar{t}_j) \rangle,
\label{B}
\end{eqnarray}
where
\begin{equation}
C_{k_ik_jq_iq_j} =( u_{q_i} \gamma_{k_i} + v_{q_i} \gamma_{k_i+q_i} )
    ( u_{q_i} \gamma_{k_j-q_i} +v_{q_i} \gamma_{k_j} ),
\end{equation}
 We find that
\begin{equation}
C_{k_ik_jq_iq_j}  =
  \frac{1}{8 \nu_{q_i}} \left[ (q_{ix} \sin k_{x}+ q_{iy} \sin k_{y} )^2
            + O (q^3, q^2 \cos^2 k) \right],
\label{C}
\end{equation}
for $k$ near the band minimum $k_0$ and for small momentum transfers $q$.
The important feature of Eq (\ref{C}) is that $C_{k_ik_jq_iq_j}$ does
not depend on the exact value of the hole momenta, $k_i$ and $k_j$,
but only on the momentum transfer $q_i$ and
the initial momentum $k$. This result requires only that 
$\gamma_q= 1  + O(q^2)$ for small
$q$ and, for $k = k_0 + (\delta k_x, \delta k_y)$,
$\gamma_k = (\delta k_x + \delta k_y)
+ O(\delta k^3)$. We can then formally rewrite
$G_{m}(k,\bar{t})$ in (\ref{Gexp2}) as
\begin{eqnarray}
G_{m}(k,\bar{t}) &=& \frac{(itz)^{2m}}{(2m)!N^m}
       \sum_{q_1 \dots q_{2m}}
        \int_0^{\bar{t}} d\bar{t}_1 \dots \int_0^{\bar{t}} d \bar{t}_{2m}
\nonumber \\
  &&  \mbox{   } \langle  T \bar{M}_{k, q_{2m}}(\bar{t}_{2m}) \cdots
    \bar{M}_{k,q_1} (\bar{t}_1)  \rangle,
\label{G2m'}
\end{eqnarray}
where
\begin{equation}
\bar{M}_{k,q} (\bar{t})=\frac{1}{\sqrt{8\nu_q}}| q_{x} \sin k_{x}   +
  q_{y} \sin k_{y} |
  [ \beta_{q}(\bar{t}) + \beta_{-q}^\dagger (\bar{t}) ].
\end{equation}
Using (\ref{Gexp}), we have
\begin{equation}
G(k,\bar{t}) = -i \theta(\bar{t})
          \langle  T \displaystyle{ e^{
   -i tz \int_0^{\bar{t}} d \bar{t}_1 \sum_q \bar{M}_{k,q} (\bar{t}_1)}}
      \rangle
\end{equation}
and hence \cite{note1}
\begin{equation}
G(k,\bar{t}) =  -i \theta(\bar{t})
        e^{ \displaystyle{- i \epsilon_k \bar{t}
                 -\phi_k (\bar{t}) }},
\label{Gres}
\end{equation}
with
\begin{eqnarray}
\epsilon_k &=& - \frac{t^2}{J}
     \frac{1}{N} \sum_q \frac{q_{x}^2 \sin^2 k_{x}+  q_{y}^2 
      \sin^2 k_{y} }{\nu_q^2},
       \nonumber  \\
\phi_k(\bar{t}) &=& -  \frac{1}{2}\left( \frac{t}{J} \right)^2
       \frac{1}{N} \sum_q  \frac{ q_x^2  \sin^2 k_{x}+ 
                        q_y^2  \sin^2 k_{y}}{\nu_q^3}
    [ (e^{-i \omega_q \bar{t}} -1) (N_q+1)+(e^{i \omega_q \bar{t}} -1)N_q ].
\label{phi}
\end{eqnarray}

Here we will only discuss properties of the hole at zero temperature.
At $T=0$, $N_q \to 0$.  
In 1D, after performing the sum (integration  for an infinite system) 
over $q$ in (\ref{phi}),
we find that the function
$\phi_{k} (\bar{t})|_{k=\pi/2} \sim
\ln(1+i\xi \tilde{t})$ here $\tilde{t} = \bar{t}/Jzs$ and the momentum cut-off
$\xi \sim O(1)$. We have approximated the spin spectrum by linear 
dispersion.
The logarithmic divergence at large times leads to
the so-called orthogonality catastrophe. The corresponding
spectral function $A(k_0,\omega)  \sim$
$\theta(\omega-\epsilon_{k_0}) (\omega-\epsilon_{k_0})^{g-1}$,
with $g=(t/J)^2/2\pi$, and there is
no quasiparticle behavior in this case.
In 2D, however, we obtain that $\phi_{k_0} (\bar{t})$ $\sim
i\tilde{t}^{-1}(e^{-i\xi \tilde{t}}-1)-\xi$.
For large $\bar{t}$ (or $\tilde{t}$),
since the first term is irrelevant, there is no
logarithmic term.
The constant term left in  $\phi_{k_0} (\bar{t})$
at the large $\bar{t}$ limit  contributes a finite quasiparticle weight and
the spectral function  $A(k,\omega)$  is then of the form
\begin{equation}
A(k,\omega) = 2\pi Z_k \delta(\omega-\epsilon_k)+A_{\rm inc} (k, \omega),
\label{Ak}
\end{equation}
where the quasiparticle weight $Z_k=\exp[-c_k (t/J)^2]$ ($c_k$ are constants).
This exponential factor
is reminiscent of the
Huang-Rhys factor  \cite{Mahan,Huang} in the usual electron-phonon problem,
indicating a polaronic behavior. 
This Huang-Rhys factor is in agreement with that obtained using 
the BdeG equation.\cite{Su}

Using (\ref{Gres}) and linear dispersion as the spin wave spectrum, we
obtain that in two dimensions the hole spectral function at $k_0$ is given by
\begin{equation}
A(k_0,\omega) = 2\pi Z \delta(\omega-\epsilon_k)
              + {\rm Re} \int_0^\infty d \bar{t}
           e^{i(\omega-\epsilon_{k_0}) \bar{t}  } Z \left[ 
     \displaystyle{ e^{i\alpha (e^{-i\xi c \bar{t}}-1)/\bar{t} }}
              -1 \right],
\label{Ak1}
\end{equation}
where $c=Jzs$ and $\alpha=(t/J)^2/(4\pi c)$.  The second term 
in (\ref{Ak1}), {\it i.e.\/} the incoherent part, is well-behaved.  
$A(k,\omega)$ is shown in Fig.\ 1.   The incoherent part is almost 
constant over a broad energy region.  In 
the results of very small cluster calculations (both
exact and SCBA) \cite{review,LM95},
many secondary peaks in $A(k,\omega)$ were 
found above the lowest quasiparticle 
one. These
secondary peaks were attributed to ``string'' resonances.  However, the  
cluster calculations also show that, when the size of the system
increases, these peaks become less prononced and,
recently, Leung and Gooding \cite{Leung} found in exact diagonalizations 
that the secondary
peaks which are well defined for a 16-site lattice 
disappear in a 32-site system and that the secondary peaks are a finite
size effect. Our results are consistent with this suggestion. If 
the secondary peaks were smeared out,the spectral 
function obtained from the small-cluster calculations would have 
the same broad feature as  that shown in Fig.\ 1.

We note that the hole  Green's function (\ref{Gres}) 
has the same form as that of
the first-order approximation in the cumulant expansion.
In fact, in the cumulant expansion \cite{Mahan}
\begin{equation}
G(k,\bar{t}) =  -i \theta(\bar{t})
        \exp \{ \sum_{n=1}^{\infty}  F_n(k,\bar{t}) \},
\end{equation}
where 
\begin{eqnarray}
&F_1(k,\bar{t})=G_1 (k,\bar{t}), \nonumber \\
&F_2(k,\bar{t})=G_2 (k,\bar{t})-\frac{1}{2!} F_1^2,  \\
&F_3(k,\bar{t})=G_3 (k,\bar{t})-F_1F_2-\frac{1}{3!} F_1^3,  \cdots.\nonumber 
\end{eqnarray}
Here $G_i$ is defined in (\ref{Gexp2}).  The first order term is given by
\begin{equation}
 F_1(k,\bar{t}) =  (tz)^2
      \sum_q  \frac{(u_q \gamma_{k-q}+ v_q \gamma_k)^2}{\omega_q}
    \left\{ i \bar{t} +
\left[ (e^{-i \omega_q \bar{t}} -1) 
(N_q+1)+(e^{i \omega_q \bar{t}} -1)N_q \right] / \omega_q \right\}.
\label{FO}
\end{equation}
Since in the long-wavelength limit the spin waves are uncorrelated for the 
hole at the band minimum, $F_i (k,\bar{t}) =0$ for $i \ge 2$
and $F_1 (k,\bar{t}) =  - i \epsilon_k \bar{t}  -\phi_k (\bar{t}) $.
So the result of the first-order approximation in the 
cumulant expansion is exact in the long-wavelength limit for 
the hole momemtum $k=k_0$.
For the usual polaron problem \cite{Mahan1,Dunn},
numerical calculations have
indicated that the cumulant expansion converges rapidly for weak and 
intermediate couplings \cite{Dunn}.  
But quite why the resummation into the exponential
like the cumulant expansion is a proper choice for the problem has not
been understood clearly.  Here for the spin polaron problem, we 
establish a connection between 
the first order approximation in the cumulant expansion and
the long-wavelength approximation for the boson excitations.

The LWA gives that the energy of the hole scales as $t^2/J$,
as seen in (\ref{phi}), 
which should be correct only in the large $J/t$ case.
In the intermediate and strong coupling regions, the BdeG study \cite{Su}
suggests that the energy should be multiplied by the Huang-Rhys factor,
in which case the bandwidth should be represented by
\begin{equation}
W/t = a \frac{t}{J} e^{- c (t/J)^2},
\label{bandwidth}
\end{equation}
where $a$ and $c$ are independent of $t$ and $J$.
To see how well this
universal functional can represent the hole bandwidth, 
we compare it with numerical results of other studies.
We show various estimates for the hole bandwidth in Fig.\ 2.
The dashed-dotted line  and ``$\star$'' correspond  to results obtained
using the variational approach by Sachdev\cite{Sachdev} and
ED on clusters of 20 sites by Poilblanc et al \cite{PZSD}.  The variational
approach is reliable in the large $J/t$ case, while most ED's
are only available for small $J/t$. 
We choose $a=2.8$ and $c=0.5$ for the functional dependence to fit 
results of other studies.  The functional dependence is shown by the solid
curve in Fig.\ 2.  It is
very close to the variational results for large $J/t$ and
to the ED results for small $J/t$, especially $J/t \ge 0.4$.
For small $J/t$,  a power law fit to
the functional dependence gives
$b+d(J/t)^{0.667}$ (the dashed curve).
The coefficients $b$ and $d$ are different from
those obtained from the numerical calculations \cite{PZSD,Leung,review}.
The form (\ref{bandwidth}) is thus only qualitatively correct 
in the small $J/t$ limit. In the region of $J/t$
between 1.0 and 2.0,  the functional dependence  describes a
smooth crossover from $t/J$ behavior
in the $J/t$ limit to  roughly $J/t$ behavior in the small $J/t$ limit.
For comparison,  the results of
SCBA (open circles) are also shown in Fig.\ 1.
The SCBA  seems to substantially underestimate values of the
bandwidth in the intermediate and the large $J/t$ cases \cite{LM}.

We would like to mention that although the energy of the hole 
$\epsilon_k$ in (\ref{phi}), which  is proportional to
$\sin^2 k_x + \sin^2 k_y$, gives rise to the hole pockets at
$k_0$ correctly, the other terms like $\cos k_x \cos k_y$,
which lead to anisotropy of the effective mass,  are of  higher
order in LWA.  So the LWA cannot be expected to give the anisotropy
quantitatively. To discuss the anisotropy of the effective mass qualitatively,
one could use the result of the first order approximation of the 
cumulant expansion, i.e., Eq.\ (\ref{FO}).

In conclusion,
we  have derived an analytical expression for
the Green's function of the hole moving in an
antiferromagnet near the band mimimum in the long wavelength limit.
The Green's function clearly indicates that the infrared divergence is
eliminated in two dimensions so that the quasiparticle weight is finite.
It also suggests that the hole motion has a polaronic nature
for intermediate and large $J/t$.
We have shown that the cumulant expansion is a good choice for studying
the hole motion in the weak and intermediate coupling cases, with
the first-order approximation equivalent to the long-wavelength 
approximation at the band minimum of the hole.  This should be 
complementary to the self-consistent Born approximation 
which is better for small $J/t$ limit.

One of the authors (YML) acknowledges
support from  the EPSRC of the United Kingdom
under grant No.\ GRK42233 and from MURST/British Council under grant
No.\ Rom/889/92/47.


\begin{figure}
\caption{  The hole spectral function $A(k_0, \omega)$ as a function of 
$\omega$.  We take  $\xi=\pi$ and  $J/t=1.0$. 
$\omega$  is in unit of $Jzs$.}
\end{figure}

\begin{figure}
\caption{ The hole bandwidth $W/t$   as a
 function of $J/t$ for a hole moving in a 2D antiferromagnetic background.
 The functional dependence  
$W/t = 2.8 \frac{t}{J} e^{- 0.5 (t/J)^2}$ (see the text) 
is shown by the solid curve,
along with the results from  the
exact diagonalization calculations on 20 sites (``$\star$'')
\protect\onlinecite{PZSD},
from the self-consistent Born approximation on
a cluster of $16 \times 16$ (open circles),\protect\cite{Martinez}
and using a variational approach.\protect\cite{Sachdev}
The dashed curve, which is proportional
to $(J/t)^{0.677}$, is the best fit to  the solid curve for small $J/t$.}
\label{Fig:mass}
\end{figure}


\begin{references}
\bibitem{note1}  In deriving Eq.\ (\protect\ref{Gres}) we use the relation
     $\langle T \exp \{\int_0^t  d t \bar{M} (t)\} \rangle$
   $=\exp \{\frac{1}{2} \int_0^t  d t_1 \int_0^t  d t_2 $
  $ \langle T  \bar{M} (t_1) \bar{M} (t_2) \rangle \}$,
  where $\bar{M}$ is
     a linear combination of boson operators $\beta_q^\dagger$ and
    $\beta_q$.  Equation (\protect\ref{Gres})
        can also be obtained from the cumulant expansion
       from the usual polaron problem, see \onlinecite{Mahan}.
\end{references}
\end{document}